\documentclass[twocolumn,english,aps,pra]{revtex4}
\usepackage[T1]{fontenc}
\usepackage[latin9]{inputenc}
\usepackage{amsmath}
\usepackage{graphicx}

\makeatletter
\@ifundefined{textcolor}{}
{%
 \definecolor{BLACK}{gray}{0}
 \definecolor{WHITE}{gray}{1}
 \definecolor{RED}{rgb}{1,0,0}
 \definecolor{GREEN}{rgb}{0,1,0}
 \definecolor{BLUE}{rgb}{0,0,1}
 \definecolor{CYAN}{cmyk}{1,0,0,0}
 \definecolor{MAGENTA}{cmyk}{0,1,0,0}
 \definecolor{YELLOW}{cmyk}{0,0,1,0}
}

\makeatother

\begin{document}
\title{Artificial Light Harvesting by Dimerized M\"{o}bius Ring}

\author{Lei Xu}

\address{Beijing Computational Science Research Center, Beijing 100084, China}

\author{Z. R. Gong}

\address{School of Physical Sciences and Technology, Shenzhen University,
Shenzhen 518060, China}

\author{Ming-Jie Tao}

\address{Department of Physics, Applied Optics Beijing Area Major Laboratory,
Beijing Normal University, Beijing 100875, China}

\author{Qing Ai}
\email{aiqing@bnu.edu.cn}


\address{Department of Physics, Applied Optics Beijing Area Major Laboratory,
Beijing Normal University, Beijing 100875, China}
\begin{abstract}
We theoretically study artificial light harvesting by a dimerized
M\"{o}bius ring. When the donors in the ring are dimerized, the energies
of the donor ring are splitted into two sub-bands. Because of the
nontrivial M\"{o}bius boundary condition, both the photon and acceptor
are coupled to all collective-excitation modes in the donor ring.
Therefore, the quantum dynamics in the light harvesting are subtly
influenced by the dimerization in the M\"{o}bius ring. It is discovered
that energy transfer is more efficient in a dimerized ring than that
in an equally-spaced ring. This discovery is also confirmed by the
calculation with the perturbation theory, which is equivalent to the
Wigner-Weisskopf approximation. Our findings may be beneficial to
the optimal design of artificial light harvesting.
\end{abstract}
\maketitle

\section{Introduction}

Photosynthesis is the main resource of energy supply for living beings
on earth. Therein, efficient light harvesting process, which delivers
the captured photon energy to the reaction center, plays a crucial
role in natural photosynthesis \cite{Blankenship02,May04,Amerongen00,Fleming08}.
Because a series of experimental and theoretical explorations
\cite{Engel07,Lee07,Collini10,Hildner13,Ishizaki09} have demonstrated that quantum coherent phenomena might exist and
even optimize the natural photosynthesis, much effort has been made
to reveal the effect of quantum coherence on efficient light harvesting
\cite{Cheng06,Chin13,Ai13,Sarovar10,Liao10,Mohseni08,Qin14,Tao16}.

The researches on optimal geometries for efficient energy transfer
are rare \cite{Ai13,Wu13,Hoyer12,Yang10,Dong12}. Wu \textit{et al}.
demonstrated the trapping-free mechanism for efficient light harvesting
in a star-like artificial system \cite{Wu13}. Hoyer \textit{et al}.
showed ratchet effect for quantum coherent energy transfer in a one-dimensional
system \cite{Hoyer12}. In 2013, one of the authors Q.A. and his collaborators
elucidated that clustered geometries utilize exciton delocalization
and energy matching condition to optimize energy transfer in a generic
tetramer model \cite{Ai13} as well as in Fenna-Matthews-Olson complex
\cite{Sarovar11,Nalbach11}. However, photosynthesis with ring-shape
geometry is more frequently observed in natural photosynthetic complexes,
e.g. LH1 and LH2 \cite{Cheng06,Olaya-Castro08}. This observation
inspired Yang \textit{et al}. to prove that symmetry breaking in B850
ring of LH2 complex boosts efficient inter-complex energy transfer
\cite{Yang10}. Dong \textit{et al}. further proposed that a perfect
donor ring for artificial light harvesting makes full use of collective
excitation and dark state to enhance the energy transfer efficiency
\cite{Dong12}.

Mathematically speaking, the rings as in LH1 and LH2 are topologically
trivial, since both the photon and acceptor are only coupled to the
zero-momentum collective-excitation mode regardless of the dimerization
as shown in Appendix~\ref{app:RingPBC}. On the other hand, M\"{o}bius
strips \cite{Heilbronner64,Ajami03,Yoneda14,Zhao09} manifest novel
physical properties and can be used to fabricate novel devices and
materials \cite{Balzani08}, e.g. topological insulators and negative-index
metamaterials \cite{Guo09,Chang10,Fang16}. In these M\"{o}bius strips,
the electrons in the ring experience different local effective fields
at different positions due to the topologically non-trivial boundary
condition. This observation enlightens us on the investigation of
the M\"{o}bius strips in artificial light harvesting. When the donors
in the ring are dimerized, there are two energy sub-bands for collective
excitation modes. Due to the M\"{o}bius boundary condition, both
the photon and acceptor interact with all collective excitations in
the ring. This is in remarkable contrast to the case in Ref.~\cite{Dong12},
where they are only coupled to single collective excitation mode.

This paper is organized as follows: the donor ring with M\"{o}bius
boundary condition in introduced for light harvesting in the next
section. Then, in Sec.~\ref{numerical}, the energy transfer efficiency
is numerically simulated for various of parameter regimes. Finally,
the main results are summarized in the Conclusion part. In Appendix~\ref{app:Diagonalization},
a brief description of diagonalizing a dimerized ring with M\"{o}bius
boundary condition is given. In Appendix~\ref{app:RingPBC}, we prove
that both the photon and acceptor only interact with one of the collective
excitation modes in the ring with periodical boundary condition, no
matter whether the dimerization exists in the ring or not. In Appendix~\ref{app:WWapp},
we present a detailed perturbation theory for describing energy transfer
in a ring with M\"{o}bius boundary condition.

\section{Dimerized Ring with M\"{o}bius Boundary Condition}

\label{model}

\begin{figure}
\includegraphics[bb=0bp 0bp 360bp 358bp,clip,width=6cm]{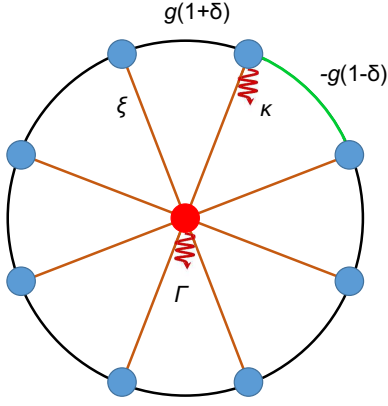}
\caption{(color online) Schematic of light harvesting by dimerized ring with
M\"{o}bius boundary condition. $N$ donors form a dimerized ring
with equal site energy. The green curve on the ring labels the M\"{o}bius
boundary condition. Due to dimerization, there are two different kinds
of alternative couplings between nearest neighbors, i.e. $g(1\pm\delta)$.
An acceptor is located in the center of ring with site energy $\varepsilon_{A}$.
All donors are uniformly coupled to the acceptor with strength $\xi$.
The fluorescence rate of the donor is $\kappa$, while the charge
separation rate on the acceptor is $\Gamma$.}
\end{figure}

\begin{figure}
\includegraphics[bb=90bp 235bp 490bp 540bp, clip, width=9cm]{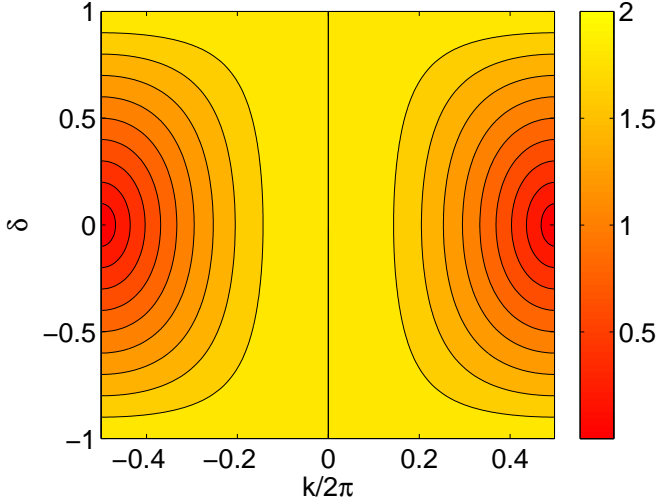}
\caption{(color online) Energy spectrum $\varepsilon_{k}$ of $A_{k}$ modes
vs dimerization $\delta$ with $N=800$, and $g/\xi=1$, and $\xi=10\mathrm{ps}^{-1}$.
For any given $\delta$, $\varepsilon_{k}$ varies in the range $[2g|\delta|,2g]$.\label{fig:Ek}}
\end{figure}

In this paper, we consider the light harvesting in a Peierls distorted
chain with M\"{o}bius boundary condition \cite{Peierls55,Shen13}.
The quantum dynamics of the whole system is governed by the Hamiltonian
\begin{eqnarray}
H & = & \omega b^{\dagger}b+\varepsilon_{A}A^{\dagger}A+H_{\mathrm{PM}}+\sum_{j=1}^{N}(\xi d_{j}^{\dagger}A+Jd_{j}^{\dagger}b)+\mathrm{h.c.},\nonumber \\
\end{eqnarray}
where $b^{\dagger}$($b$) is the creation (annihilation) operator
of photon with frequency $\omega$, $A^{\dagger}$($A$) is the creation
(annihilation) operator of excitation at the acceptor with site energy
$\varepsilon_{A}$, $d_{j}^{\dagger}$($d_{j}$) is the creation (annihilation)
operator of an excitation at $j$th donor, and the Peierls distorted
ring with M\"{o}bius boundary condition is described by
\begin{equation}
H_{\mathrm{PM}}=\sum_{j=1}^{N-1}g[1-(-1)^{j}\delta]d_{j+1}^{\dagger}d_{j}-g[1-(-1)^{N}\delta]d_{1}^{\dagger}d_{N}+\mathrm{h.c.}
\end{equation}
with $g$ being coupling constant between nearest neighbors and $\delta$
being dimerization constant. Notice that the minus sign before the
second term on the r.h.s. indicates the M\"{o}bius boundary condition
\cite{Shen13}, and the site energies of donors are homogeneous and
chosen as the zero point of energy. Here we assume that the photon
and acceptor are coupled to all donors with equal coupling strength
$J$ and $\xi$, respectively.

By the diagonalization method of $H_{\mathrm{PM}}$ in Appendix~\ref{app:Diagonalization},
the total Hamiltonian is rewritten as
\begin{eqnarray}
H & = & \omega b^{\dagger}b+\varepsilon_{A}A^{\dagger}A+\sum_{k}\varepsilon_{k}(A_{k}^{\dagger}A_{k}-B_{k}^{\dagger}B_{k})+H_{1}.\label{eq:H}
\end{eqnarray}
There are two energy bands in the donor ring denoted by the annihilation
(creation) operators
\begin{eqnarray}
A_{k} & = & \sum_{j=1}^{N/2}\frac{e^{-ikj}}{\sqrt{N}}(e^{i(2j-1)\frac{\pi}{N}}d_{2j-1}+e^{i(\theta_{k}+2j\frac{\pi}{N})}d_{2j}),\\
B_{k} & = & \sum_{j=1}^{N/2}\frac{e^{-ikj}}{\sqrt{N}}(e^{i(2j-1)\frac{\pi}{N}}d_{2j-1}-e^{i(\theta_{k}+2j\frac{\pi}{N})}d_{2j}),
\end{eqnarray}
 ($A_{k}^{\dagger}$ and $B_{k}^{\dagger}$) with the eigen energies
being $\pm\varepsilon_{k}$, where
\begin{eqnarray}
\varepsilon_{k} & = & 2g\sqrt{\cos^{2}(\frac{k}{2}-\frac{\pi}{N})+\delta^{2}\sin^{2}(\frac{k}{2}-\frac{\pi}{N})},
\end{eqnarray}
and the momentum
\begin{equation}
k=\frac{4\pi}{N}(0,1,2,\cdots\frac{N}{2}-1)-\pi+\frac{2\pi}{N}.
\end{equation}
The interaction Hamiltonian among the photon and acceptor and donor
ring is
\begin{align}
H_{1} & =\sum_{k}(\xi_{Ak}A_{k}^{\dagger}+\xi_{Bk}B_{k}^{\dagger})A+(J_{Ak}A_{k}^{\dagger}+J_{Bk}B_{k}^{\dagger})b+\mathrm{h.c.}\nonumber \\
 & =\sum_{k}(h_{Ak}A_{k}^{\dagger}+h_{Bk}B_{k}^{\dagger})(\xi A+Jb)+\mathrm{h.c.}
\end{align}
with the $k$-dependent factors
\begin{eqnarray}
h_{Ak} & = & \frac{1}{\sqrt{N}}\sum_{j=1}^{N/2}e^{-ikj}e^{-i(2j-1)\frac{\pi}{N}}(1+e^{i\theta_{k}}e^{i\frac{\pi}{N}}),\label{eq:Hak}\\
h_{Bk} & = & \frac{1}{\sqrt{N}}\sum_{j=1}^{N/2}e^{-ikj}e^{-i(2j-1)\frac{\pi}{N}}(1-e^{i\theta_{k}}e^{i\frac{\pi}{N}}),\label{eq:Hbk}\\
e^{i\theta_{k}} & = & \frac{g}{\varepsilon_{k}}[(1+\delta)e^{-i\frac{\pi}{N}}+(1-\delta)e^{-i(k-\frac{\pi}{N})}].
\end{eqnarray}

Before investigating the quantum dynamics of light harvesting, we
shall analyze the energy spectrum of M\"{o}bius ring for different
dimerization $\delta$. As shown in Fig.~\ref{fig:Ek}, the characteristics
of the energy spectrum vary remarkably in response to the change of
$\delta$. When the donors in the ring are equally distributed, i.e.
$\delta=0$, the energy spectrum $\varepsilon_{k}=2g\cos(k/2)$ changes
over a range $2g$. For other parameters, i.e. $0<|\delta|<1$, the
energy spectrum $\varepsilon_{k}$, which lies in the range $[2g|\delta|,2g]$,
shrinks as $|\delta|$ approaches unity. There is an energy gap between
the two sub-bands $4g|\delta|$.

Previously, M\"{o}bius strip \cite{Yoneda14,Heilbronner64,Zhao09}
was proposed to fabricate novel devices and materials \cite{Balzani08},
e.g. topological insulators and negative-index metamaterials \cite{Guo09,Fang16}.
Although it seems that the ring with M\"{o}bius boundary condition
in this paper is somewhat different from M\"{o}bius strip in Refs.~\cite{Zhao09,Guo09,Fang16},
we remark that the ring with M\"{o}bius boundary condition can be
considered as M\"{o}bius strip in the Hilbert space. Further comparison
shows that the ring with M\"{o}bius boundary condition is not evenly
twisted as the previous M\"{o}bius strip. As a result of M\"{o}bius
boundary condition, the photon and acceptor are coupled to all the
collective modes in the ring, and thus leads to different quantum
dynamics as compared to that for a ring with periodical boundary condition.

\section{Numerical Results}

\label{numerical}

\begin{figure}[t]
\includegraphics[bb=90bp 235bp 490bp 540bp, clip, width=9cm]{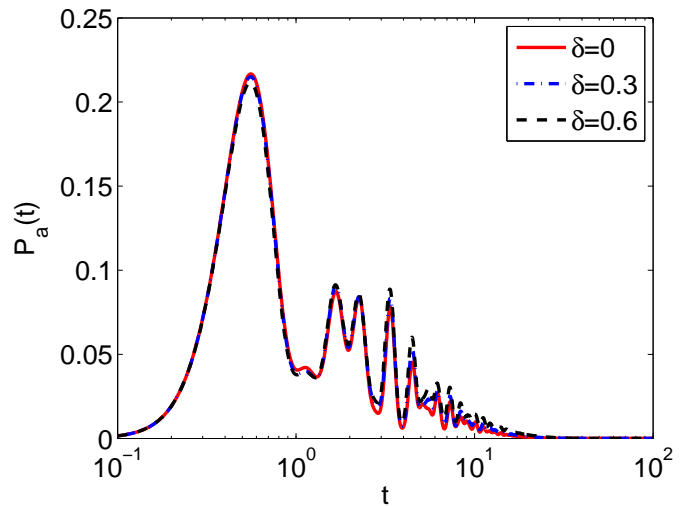}
\caption{(color online) Population dynamics of acceptor $P_{a}(t)=|\alpha_{A}|^{2}$
with M\"{o}bius ring for red solid line $\delta=0$, blue dash-dotted
line $\delta=0.3$, and black dashed line $\delta=0.6$. Other parameters
are $N=8$, $\omega/\xi=\varepsilon_{A}/\xi=-6$, $J/\xi=1$, $g/\xi=1$,
$\Gamma/\xi=0.3$ and $\kappa/\Gamma=1$ with $\xi=10\mathrm{ps}^{-1}$.
\label{fig:PaDynamics}}
\end{figure}

\begin{figure}
\includegraphics[bb=90bp 235bp 490bp 540bp, clip, width=4cm]{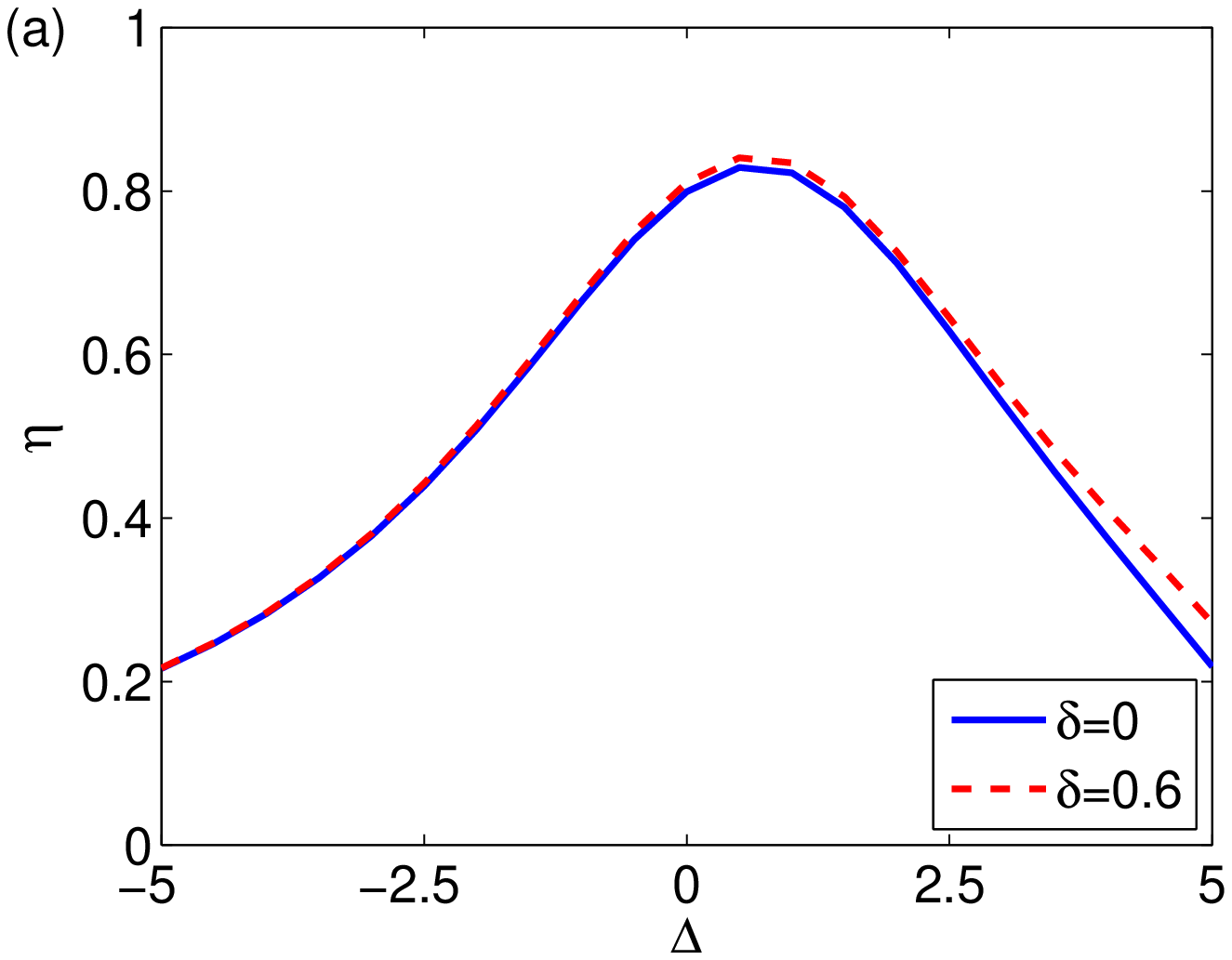}
\includegraphics[bb=90bp 235bp 490bp 540bp, clip, width=4cm]{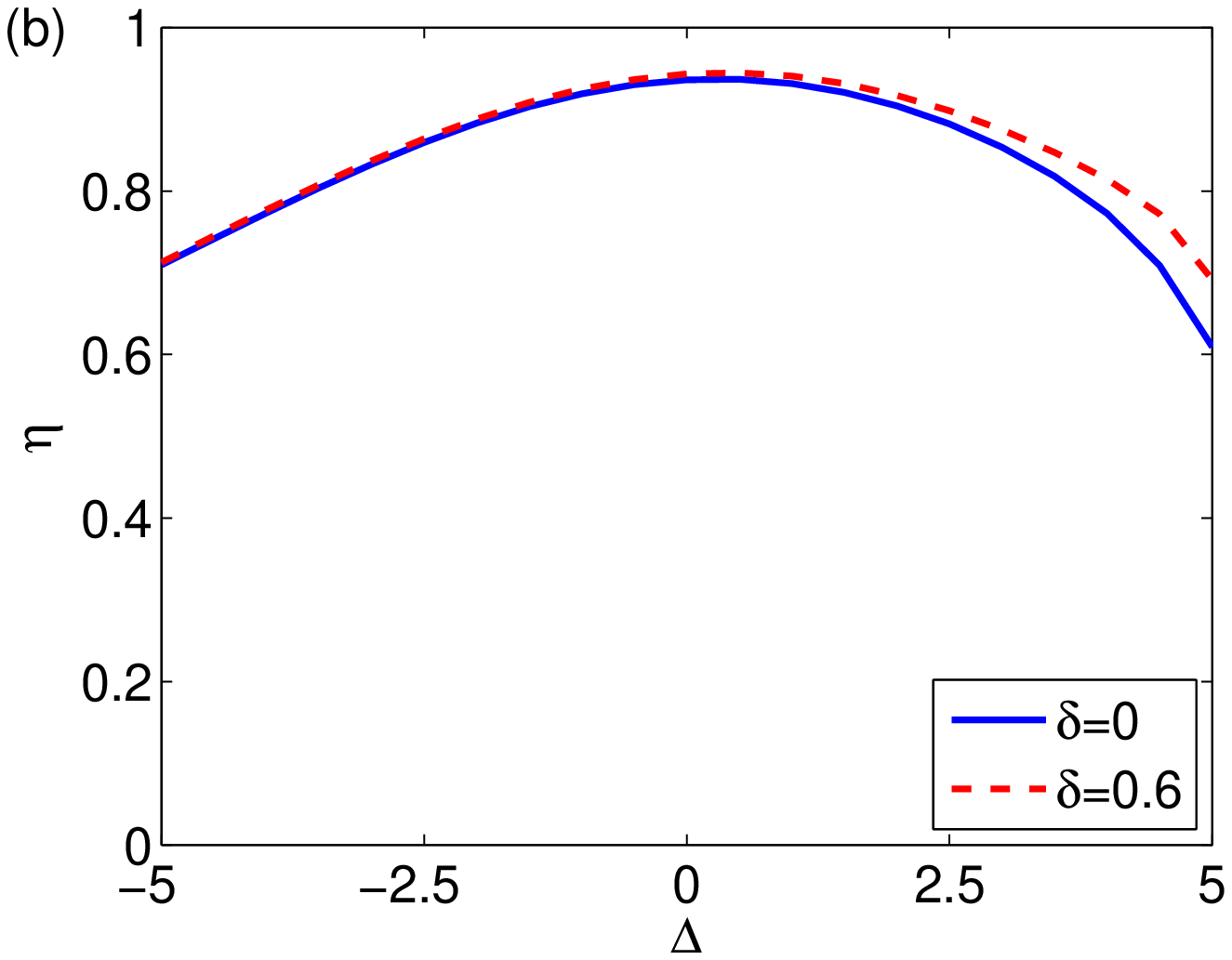}
\caption{(color online) Transfer efficiency $\eta$ vs detuning $\Delta=\omega-\varepsilon_{A}$
for (a) a large fluorescence rate $\kappa/\xi=1$; (b) a small fluorescence
rate $\kappa/\xi=0.1$. In each subfigure, we plot the efficiency
for a dimerized M\"{o}bius ring with $\delta=0.6$ (red dashed line)
and for an equally-spaced M\"{o}bius ring with $\delta=0$ (blue
solid line). Other parameters are $N=8$, $\varepsilon_{A}/\xi=-6$,
$J/\xi=1$, $g/\xi=1$, and $\Gamma/\xi=0.3$ with $\xi=10\mathrm{ps}^{-1}$.
\label{fig:EtaVsDelta}}
\end{figure}

In this section, we consider the quantum dynamics of light harvesting
by a Peierls distorted chain with M\"{o}bius boundary condition.
For an initial state $\vert\psi(0)\rangle=\vert1_{b}\rangle\equiv\vert1_{b},0_{A},0_{Ak},0_{Bk}\rangle$,
the state of the total system at any time $t$
\begin{equation}
\vert\psi(t)\rangle=\alpha_{b}\vert1_{b}\rangle+\alpha_{A}\vert1_{A}\rangle+\sum_{k}\beta_{Ak}\vert1_{Ak}\rangle+\sum_{k}\beta_{Bk}\vert1_{Bk}\rangle,
\end{equation}
with $\vert1_{A}\rangle\equiv\vert0_{b},1_{A},0_{Ak},0_{Bk}\rangle$,
$\vert1_{Ak}\rangle\equiv\vert0_{b},0_{A},1_{Ak},0_{Bk}\rangle$,
$\vert1_{Bk}\rangle\equiv\vert0_{b},0_{A},0_{Ak},1_{Bk}\rangle$,
is governed by the Schr\"{o}dinger equation
\begin{equation}
i\partial_{t}\vert\psi(t)\rangle=H\vert\psi(t)\rangle,
\end{equation}
if the system does not interact with the environment. However, an
open quantum system inevitably suffers from decoherence due to the
couplings to the environment. Generally speaking, the quantum dynamics
in the presence of decoherence is described by the master equation
instead of Schr\"{o}dinger equation. Despite this, it has been shown
that the quantum dynamics for light harvesting can be well simulated
by the quantum jump approach with a non-Hermitian Hamiltonian where
the imaginary parts in the diagonal terms represent the decoherence
processes \cite{Olaya-Castro08,Ai14}. In this case, the Hamiltonian
$H$ in Eq.~(\ref{eq:H}) is obtained by replacing the energies of
the acceptor and collective-excitation modes in the following way,
i.e.\begin{subequations}
\begin{eqnarray}
A:\varepsilon_{A} & \rightarrow & \varepsilon_{A}^{\prime}=\varepsilon_{A}-i\Gamma,\\
A_k:\varepsilon_{k} & \rightarrow & \varepsilon_{k}^{-}=\varepsilon_{k}- i\kappa,\\
B_k:-\varepsilon_{k} & \rightarrow & -\varepsilon_{k}^{+}=-(\varepsilon_{k}+ i\kappa),
\end{eqnarray}
\end{subequations}where $\kappa$ and $\Gamma$ are respectively
the fluorescence rate at the donors and the charge separation rate
at the acceptor.

It was shown \cite{Dong12} that in a composite system including a
photon, and an acceptor, and a perfect ring with periodical boundary
condition, the quantum dynamics of the total system can be effectively
modeled as the interaction of photon and acceptor with the donor's
single collective-excitation mode. Furthermore, as proven in Appendix~\ref{app:RingPBC},
both the photon and acceptor are still coupled to the same collective
mode of donor ring even when the ring is dimerized. However, the situation
is different when we explore the coherent energy transfer in a dimerized
M\"{o}bius ring. In Fig.~\ref{fig:PaDynamics}, we plot population
of acceptor $P_{a}(t)=\vert\alpha_{A}(t)\vert^{2}$ vs time for three
cases with different $\delta$. Clearly, the quantum dynamics in M\"{o}bius
ring is subtly influenced by the dimerization. In all cases, the populations
quickly rise to a maximum and then it is followed by damped oscillations.
After $t\simeq10$, all three curves converge to a similar same exponential
decay as the charge separation rates are the same for all three cases.

To quantitatively characterize the energy transfer, there is the transfer
efficiency $\eta$ defined as \cite{Caruso09,Fassioli10,BqAi12}
\begin{eqnarray}
\eta & = & 2\Gamma\int_{0}^{\infty}\left|\alpha_{A}(t)\right|^{2}dt.
\end{eqnarray}
In Fig.~\ref{fig:EtaVsDelta}(a), we investigate the dependence of
transfer efficiency on the detuning $\Delta=\omega-\varepsilon_{A}$
for a fast fluorescence decay $\kappa/\xi=1$. For an evenly-distributed
donor ring, i.e. $\delta=0$, the transfer efficiency reaches maximum
near the resonance, i.e. $\Delta=0$, which is slightly different
from that discovered in Ref.~\cite{Dong12}. As the photon frequency
deviates from the resonance, the transfer efficiency quickly drops.
When we further investigate the case for a dimerized ring with $\delta=0.6$,
the transfer efficiency has been increased over the whole range of
photon frequency, especially in the blue-detuned regime. The same characteristics
has also been observed in the case with a slow fluorescence decay
$\kappa/\xi=0.1$ as shown in Fig.~\ref{fig:EtaVsDelta}(b).

\begin{figure}
\includegraphics[bb=90bp 235bp 490bp 540bp, clip, width=9cm]{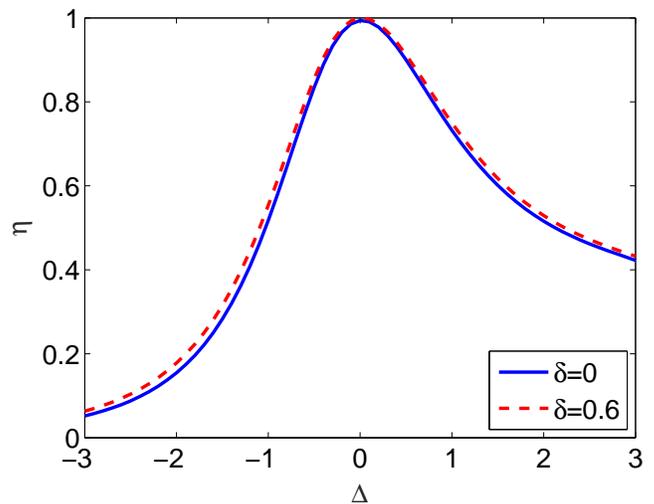}

\caption{(color online) Transfer efficiency $\eta$ vs detuning $\Delta=\omega-\varepsilon_{A}$
by the perturbation theory. The parameters are the same as in Fig.~\ref{fig:EtaVsDelta}(b).
\label{fig:EtaVsDelta-PerturbTheo}}
\end{figure}

To validate the above numerical simulation, we also calculate the
quantum dynamics and light-harvesting efficiency by the perturbation
theory. The details are presented in Appendix~\ref{app:WWapp}. In
Ref.~\cite{Wang74}, it has been proven that the Wigner-Weisskopf
approximation is equivalent to the perturbation theory in the study
of an excited state coupled to a continuum. Here we generalize the perturbation theory to the investigation of coherent energy transfer between few bodies via a continuum. In Fig.~\ref{fig:EtaVsDelta-PerturbTheo},
the energy transfer efficiency of a dimerized ring is generally larger
than that of an equally-spaced ring. This result is qualitatively consistent with Fig.~\ref{fig:EtaVsDelta}(a).

\begin{figure}
\includegraphics[bb=90bp 235bp 490bp 540bp, clip, width=4cm]{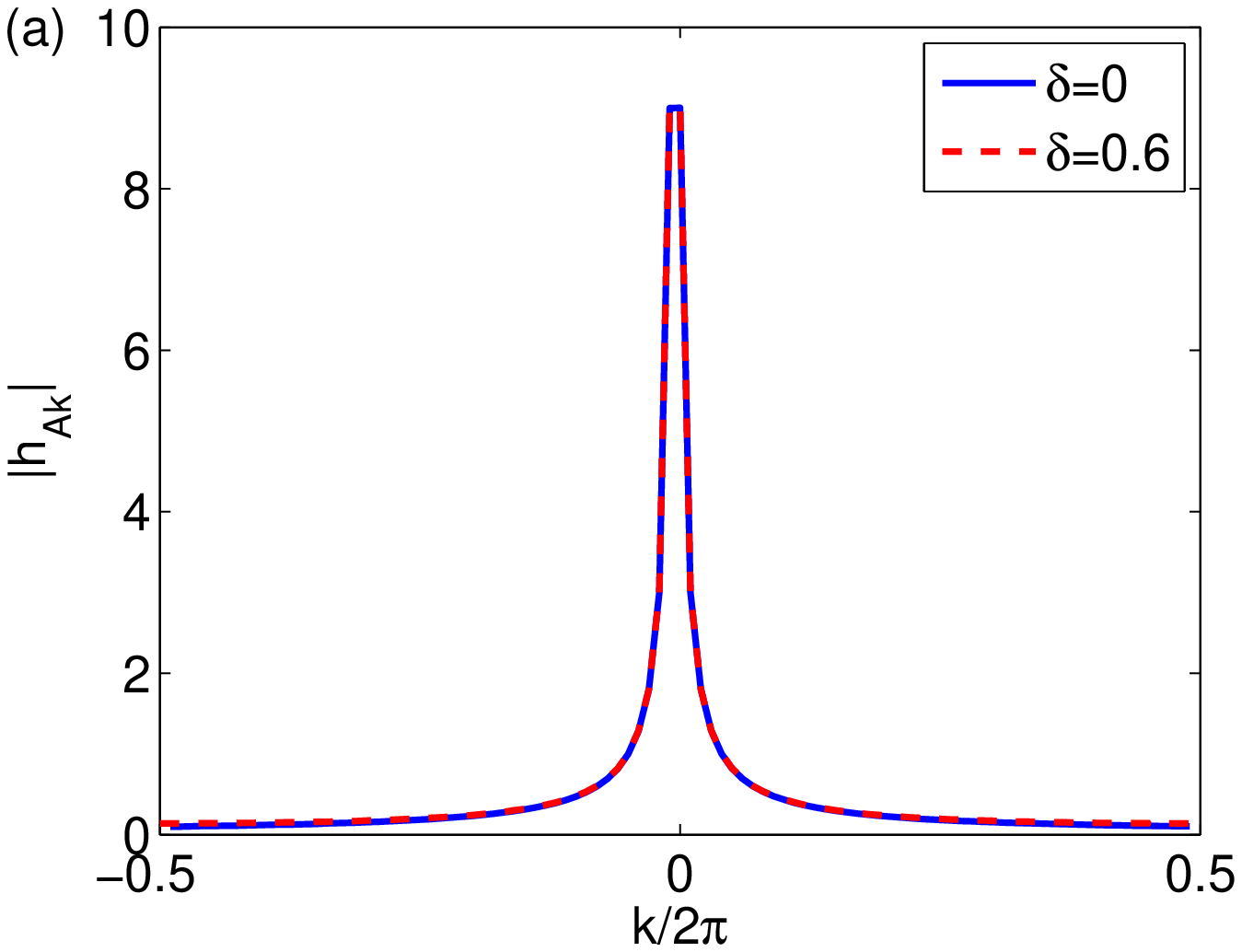}
\includegraphics[bb=90bp 235bp 490bp 540bp, clip, width=4cm]{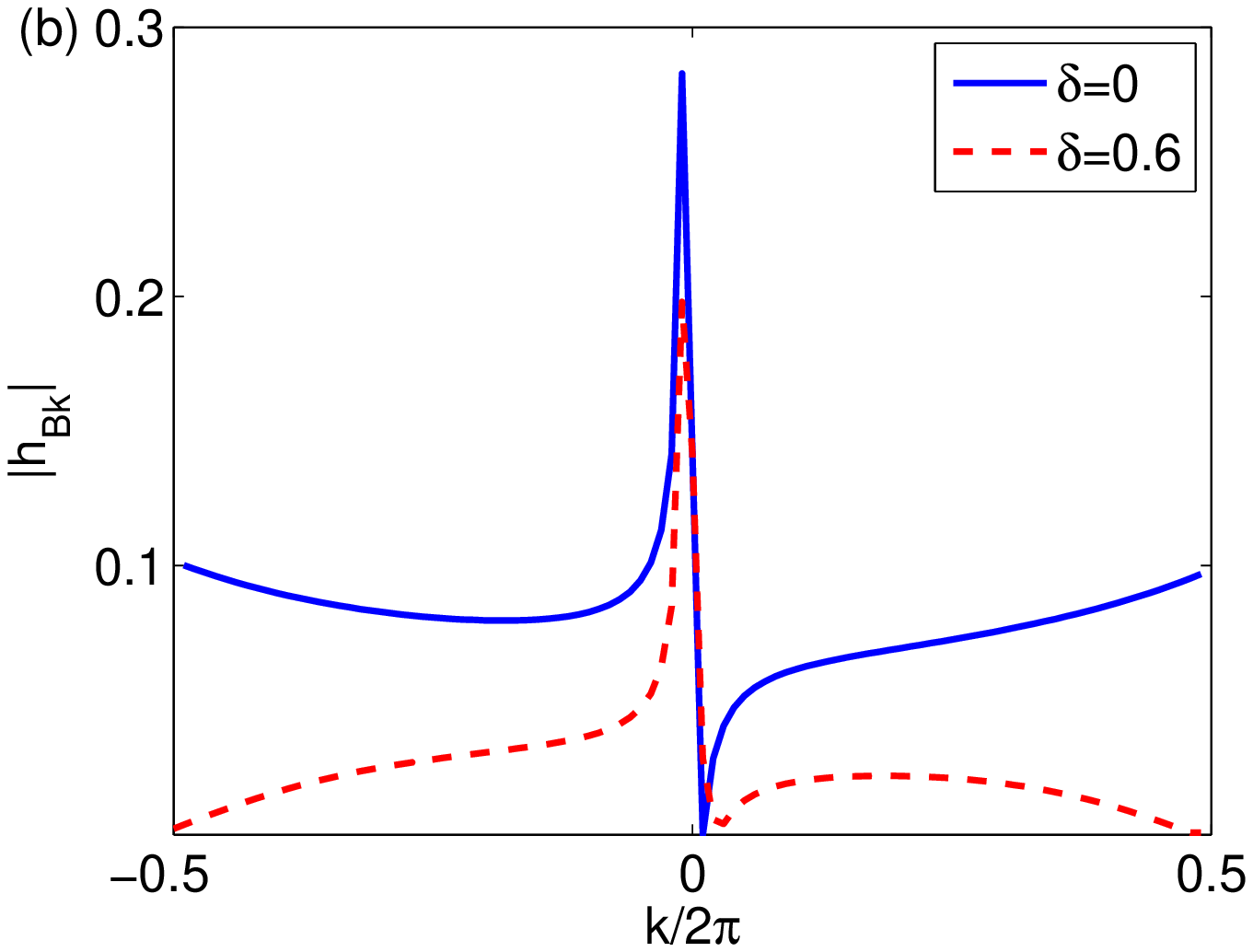}
\caption{(color online) Coupling constants (a) $|h_{Ak}|$ and (b) $|h_{Bk}|$
vs momentum $k$ for dimerization $\delta=0$ (blue solid line) and
$\delta=0.6$ (red dashed line). Other parameters are $N=200$ and
$g/\xi=1$. \label{fig:Couplings}}
\end{figure}

In order to explore the underlying physical mechanism, we turn to
coupling constants $|h_{Ak}|$ and $|h_{Bk}|$ as shown in Fig.~\ref{fig:Couplings}.
Although the coupling constants $|h_{Ak}|$ for $\delta=0$ almost
coincide with the ones for $\delta=0.6$, the coupling constants $|h_{Bk}|$
are significantly suppressed when the dimerization occurs. Furthermore,
because $|h_{Ak}|$ are generally larger than $|h_{Bk}|$ by an order,
the transfer efficiency is slightly tuned by the dimerization.

\section{Conclusion}

In this paper, we investigate the quantum dynamics of light harvesting
in a Peierls distorted ring with M\"{o}bius boundary condition. Due
to the nontrivial M\"{o}bius topology, there are two energy bands
for the excitation in the ring when the donors in the ring are dimerized.
Because both the photon and acceptor interact with all collective-excitation
modes in the M\"{o}bius ring, the quantum dynamics and thus the efficiency
for light harvesting is effectively influenced by the presence of
dimerization. By numerical simulations, we show that when the donors
in the M\"{o}bius ring are dimerized, the energy transfer is generally
optimal for a wide range of photon frequencies. Our discoveries together
with previous findings \cite{Rey13,Ai13,Wu13,Yang10,Dong12} may be
beneficial for the future design of optimal artificial light harvesting.

In addition, we also remark that
in fact change of protein environment inevitably leads to
static and dynamic disorders in the transition energies of chlorophylls in
natural photosynthesis \cite{Dong16,Pajusalu15,Sarovar11}.
In Ref.~\cite{Dong12},
assuming that all site energies of donors are homogeneous, the absorbed
photon energy utilizes dark-state mechanism to effectively avoid fluorescence
loss via the donors. However, this mechanism might not work well when
the static and dynamic disorders take place.
\begin{acknowledgments}
We thank stimulating discussions with C. P. Sun and H. Dong. This
work was supported by the National Natural Science Foundation of China
(Grant No.~11121403 and No.~11534002 and No.~11505007), the National
973 program (Grant No.~2014CB921403 and No.~2012CB922104), and the
Open Research Fund Program of the State Key Laboratory of Low-Dimensional
Quantum Physics, Tsinghua University Grant No.~KF201502.

\appendix
\end{acknowledgments}

\section{Diagonalization of M\"{o}bius Hamiltonian}

\label{app:Diagonalization}

The dimerized ring with M\"{o}bius boundary condition is described
by
\begin{equation}
H_{\mathrm{PM}}=\sum_{j=1}^{N-1}g[1-(-1)^{j}\delta]d_{j+1}^{\dagger}d_{j}-g[1-(-1)^{N}\delta]d_{1}^{\dagger}d_{N}+\mathrm{h.c.},
\end{equation}
where $\vert\delta\vert\leq1$ is a dimensionless dimerization constant.
By applying a unitary transformation
\begin{equation}
U=\sum_{j}e^{ij\frac{\pi}{N}}d_{j}^{\dagger}d_{j},
\end{equation}
the Hamiltonian reads
\begin{align}
 & UH_{\mathrm{PM}}U^{\dagger}\nonumber \\
= & \sum_{j=1}^{N-1}g[1-(-1)^{j}\delta]e^{i\frac{\pi}{N}}d_{j+1}^{\dagger}d_{j}\nonumber \\
 & -g[1-(-1)^{N}\delta]e^{-i(N-1)\frac{\pi}{N}}d_{1}^{\dagger}d_{N}+\mathrm{h.c.}\nonumber \\
= & \sum_{j=1}^{N}g_{0}[1-(-1)^{j}\delta]d_{j+1}^{\dagger}d_{j}+\mathrm{h.c.}\nonumber \\
= & \sum_{j=1}^{N/2}[g_{0}(1-\delta)d_{2j+1}^{\dagger}d_{2j}+g_{0}(1+\delta)d_{2j}^{\dagger}d_{2j-1}]+\mathrm{h.c.},
\end{align}
where
\begin{equation}
g_{0}=ge^{i\frac{\pi}{N}}.
\end{equation}
After the unitary transformation, the M\"{o}bius boundary condition
has been canceled.

By defining fermion operators \cite{Huo08}\begin{subequations}
\begin{eqnarray}
\alpha_{k} & = & \frac{1}{\sqrt{N}}\sum_{j=1}^{N/2}e^{-ikj}(d_{2j-1}+e^{i\theta_{k}}d_{2j}),\\
\beta_{k} & = & \frac{1}{\sqrt{N}}\sum_{j=1}^{N/2}e^{-ikj}(d_{2j-1}-e^{i\theta_{k}}d_{2j}),
\end{eqnarray}
\end{subequations} with inverse transformation\begin{subequations}
\begin{eqnarray}
d_{2j-1} & = & \frac{1}{\sqrt{N}}\sum_{j=1}^{N/2}e^{ikj}(\alpha_{k}+\beta_{k}),\\
d_{2j} & = & \frac{1}{\sqrt{N}}\sum_{j=1}^{N/2}e^{ikj-i\theta_{k}}(\alpha_{k}-\beta_{k}),
\end{eqnarray}
\end{subequations} where the momentum is
\begin{eqnarray}
k & = & \frac{4\pi}{N}(0,1,2,\cdots\frac{N}{2}-1)-\pi+\frac{2\pi}{N},\\
e^{i\theta_{k}} & = & \frac{g}{\varepsilon_{k}}[(1+\delta)e^{-i\frac{\pi}{N}}+(1-\delta)e^{-i(k-\frac{\pi}{N})}],\label{eq:ThetaK}\\
\varepsilon_{k} & = & 2g\sqrt{\cos^{2}(\frac{k}{2}-\frac{\pi}{N})+\delta^{2}\sin^{2}(\frac{k}{2}-\frac{\pi}{N})}\label{eq:Ek}
\end{eqnarray}
is the eigen energy, $N$ is an even number,
\begin{equation}
UH_{\mathrm{PM}}U^{\dagger}=\sum_{k}\varepsilon_{k}(\alpha_{k}^{\dagger}\alpha_{k}-\beta_{k}^{\dagger}\beta_{k}).
\end{equation}
Here the momentum is chosen in such way that $\cos(\frac{k}{2}-\frac{\pi}{N})\geq0$
for $\delta=0$ and thus the eigen energies $\varepsilon_{k}=2g\cos(\frac{k}{2}-\frac{\pi}{N})$.
Therefore, the original Hamiltonian can be diagonalized as
\begin{equation}
H_{\mathrm{PM}}=\sum_{k}\varepsilon_{k}(A_{k}^{\dagger}A_{k}-B_{k}^{\dagger}B_{k}),\label{eq:H_PDC}
\end{equation}
where the annihilation operators of collective excitation modes of
the two bands are\begin{subequations}
\begin{eqnarray}
A_{k} & = & \alpha_{k}U\\
 & = & \sum_{j=1}^{N/2}\frac{e^{-ikj}}{\sqrt{N}}(e^{i(2j-1)\frac{\pi}{N}}d_{2j-1}+e^{i(\theta_{k}+2j\frac{\pi}{N})}d_{2j}),\nonumber \\
B_{k} & = & \beta_{k}U\\
 & = & \sum_{j=1}^{N/2}\frac{e^{-ikj}}{\sqrt{N}}(e^{i(2j-1)\frac{\pi}{N}}d_{2j-1}-e^{i(\theta_{k}+2j\frac{\pi}{N})}d_{2j}).\nonumber
\end{eqnarray}
\end{subequations} According to Eq.~(\ref{eq:H_PDC}), there are
two sub-bands in the dimerized chain, i.e. $\varepsilon_{k}$ and
$-\varepsilon_{k}$. For the upper band, there is a minimum $2\vert g\delta\vert$
at $k=\pi+\frac{2\pi}{N}$, while for the lower band, there is a maximum
$-2\vert g\delta\vert$ also at $k=\pi+\frac{2\pi}{N}$. Therefore,
there is an energy gap between the two sub-bands $4\vert g\delta\vert$
as long as the ring is dimerized.

\section{Light Harvesting by Ring with Periodical Boundary Condition}

\label{app:RingPBC}

The Hamiltonian for a dimerized ring with periodical boundary condition
reads
\begin{eqnarray}
H & = & \omega b^{\dagger}b+\varepsilon_{A}A^{\dagger}A+\sum_{j}g[1-(-1)^{j}\delta]d_{j}^{\dagger}d_{j+1}\nonumber \\
 &  & +\sum_{j}(Jd_{j}^{\dagger}b+\xi d_{j}^{\dagger}A)+\mathrm{h.c.}
\end{eqnarray}

We define two sets of collective-excitation operators \begin{subequations}
\begin{eqnarray}
\alpha_{k} & = & \frac{1}{\sqrt{N}}\sum_{j=1}^{N/2}e^{-ikj}(d_{2j-1}-e^{i\theta_{k}}d_{2j}),\\
\beta_{k} & = & \frac{1}{\sqrt{N}}\sum_{j=1}^{N/2}e^{-ikj}(d_{2j-1}+e^{i\theta_{k}}d_{2j}),
\end{eqnarray}
\end{subequations} and the reverse transformation is\begin{subequations}
\begin{eqnarray}
d_{2j-1} & = & \frac{1}{\sqrt{N}}\sum_{j=1}^{N/2}e^{-ikj}(\alpha_{k}+\beta_{k}),\\
d_{2j} & = & \frac{1}{\sqrt{N}}\sum_{j=1}^{N/2}e^{-ikj-i\theta_{k}}(\beta_{k}-\alpha_{k}),
\end{eqnarray}
\end{subequations} where $k=4\pi(n-1)/N$, $n=1,2,...,N/2$, and
\begin{eqnarray}
e^{i\theta_{k}} & = & [(1+\delta)+(1-\delta)e^{-ik}]g/\varepsilon_{k},\\
\varepsilon_{k} & = & 2g\sqrt{\cos^{2}(k/2)+\delta^{2}\sin^{2}(k/2)}.
\end{eqnarray}

Thus, the Hamiltonian is transformed as
\begin{eqnarray}
H & = & \omega b^{\dagger}b+\varepsilon_{A}A^{\dagger}A+\sum_{k}\varepsilon_{k}(\beta_{k}^{\dagger}\beta_{k}-\alpha_{k}^{\dagger}\alpha_{k})\nonumber \\
 &  & +\sqrt{N}\beta_{0}^{\dagger}(Jb+\xi A)+\mathrm{h.c.}
\end{eqnarray}
Since only the zero-momentum mode
\begin{eqnarray}
\beta_{0} & = & \frac{1}{\sqrt{N}}\sum_{j=1}^{N}d_{j}
\end{eqnarray}
 with eigen energy $\varepsilon_{0}=2g$ is coupled to the photon
and acceptor, the effective Hamiltonian is further simplified as
\begin{eqnarray}
H_{\mathrm{eff}} & = & \omega b^{\dagger}b+2g\beta_{0}^{\dagger}\beta_{0}+\varepsilon_{A}A^{\dagger}A\nonumber \\
 &  & +\sqrt{N}\beta_{0}^{\dagger}(Jb+\xi A)+\mathrm{h.c.}
\end{eqnarray}
In conclusion, for a ring with periodical boundary condition, both
the photon and acceptor are coupled to the same collective-excitation
mode irrespective of the dimerization in the ring. In other words,
we have proven that the quantum dynamics of light harvesting by a
ring with periodical boundary condition is not affected by the dimerization.

\begin{widetext}

\section{Equivalence of Present Theory and Wigner-Weisskopf Approximation}

\label{app:WWapp}

The quantum dynamics in light harvesting is governed by the Schr\"{o}dinger
equation, which is equivalent to a set of coupled differential equations.
Formally, it can be solved by Laplace transformation. However, due
to the presence of branch cut, the exact solution may not be easily
obtained. Generally speaking, it can be solved by the Wigner-Weisskopf
approximation \cite{Weisskopf30,Knight72,Lee73}. In Ref.~\cite{Wang74},
it has been proven that the Wigner-Weisskopf approximation is equivalent
to the perturbation theory in the study of excited state of a few-body
system coupled to a continuum. Therefore, we make use of the perturbation
theory to investigate the quantum dynamics of light-harvesting in
a dimerized M\"{o}bius chain.

For an initial state $\vert\psi(0)\rangle=\vert1_{b}\rangle\equiv\vert1_{b},0_{A},0_{Ak},0_{Bk}\rangle$,
the state of system at any time $t$
\begin{equation}
\vert\psi(t)\rangle=\alpha_{b}(t)\vert1_{b}\rangle+\alpha_{A}(t)\vert1_{A}\rangle+\sum_{k}\beta_{Ak}(t)\vert1_{Ak}\rangle+\sum_{k}\beta_{Bk}(t)\vert1_{Bk}\rangle,
\end{equation}
with $\vert1_{A}\rangle\equiv\vert0_{b},1_{A},0_{Ak},0_{Bk}\rangle$,
$\vert1_{Ak}\rangle\equiv\vert0_{b},0_{A},1_{Ak},0_{Bk}\rangle$,
$\vert1_{Bk}\rangle\equiv\vert0_{b},0_{A},0_{Ak},1_{Bk}\rangle$,
is governed by the Schr\"{o}dinger equation
\begin{equation}
i\partial_{t}\vert\psi(t)\rangle=H\vert\psi(t)\rangle.
\end{equation}
We obtain a set of equations for the coefficients, i.e.\begin{subequations}
\begin{eqnarray}
i\dot{\alpha}_{b} & = & \omega\alpha_{b}+\epsilon\sum_{k}J_{Ak}^{*}\beta_{Ak}+\epsilon\sum_{k}J_{Bk}^{*}\beta_{Bk},\\
i\dot{\alpha}_{A} & = & \varepsilon_{A}^{\prime}\alpha_{A}+\epsilon\sum_{k}\xi_{Ak}^{*}\beta_{Ak}+\epsilon\sum_{k}\xi_{Bk}^{*}\beta_{Bk},\\
i\dot{\beta}_{Ak} & = & \varepsilon_{k}^{-}\beta_{Ak}+\epsilon\xi_{Ak}\alpha_{A}+\epsilon J_{Ak}\alpha_{b},\\
i\dot{\beta}_{Bk} & = & -\varepsilon_{k}^{+}\beta_{Bk}+\epsilon\xi_{Bk}\alpha_{A}+\epsilon J_{Bk}\alpha_{b},
\end{eqnarray}
\end{subequations}where the parameter $\epsilon$ is introduced to
keep track of the orders of perturbation.

By introducing renormalized frequencies, i.e.\begin{subequations}\label{eq:RenorFre}
\begin{eqnarray}
\Omega_{b} & = & \omega+\epsilon\Omega_{b}^{(1)}+\epsilon^{2}\Omega_{b}^{(2)}+\epsilon^{3}\Omega_{b}^{(3)}+\ldots,\\
\Omega_{A} & = & \varepsilon_{A}^{\prime}+\epsilon\Omega_{A}^{(1)}+\epsilon^{2}\Omega_{A}^{(2)}+\epsilon^{3}\Omega_{A}^{(3)}+\ldots,\\
\Omega_{Ak} & = & \varepsilon_{k}^{-}+\epsilon\Omega_{Ak}^{(1)}+\epsilon^{2}\Omega_{Ak}^{(2)}+\epsilon^{3}\Omega_{Ak}^{(3)}+\ldots,\\
\Omega_{Bk} & = & -\varepsilon_{k}^{+}+\epsilon\Omega_{Bk}^{(1)}+\epsilon^{2}\Omega_{Bk}^{(2)}+\epsilon^{3}\Omega_{Bk}^{(3)}+\ldots,
\end{eqnarray}
\end{subequations}we can define dimensionless times
\begin{eqnarray}
\tau_{b} & = & \Omega_{b}t,\:\tau_{A}=\Omega_{A}t,\:\tau_{Ak}=\Omega_{Ak}t,\:\tau_{Bk}=\Omega_{Bk}t,
\end{eqnarray}
and re-express the above differential equations as\begin{subequations}\label{eq:Diff}
\begin{eqnarray}
i\Omega_{b}\frac{\partial\alpha_{b}}{\partial\tau_{b}} & = & \omega\alpha_{b}+\epsilon\sum_{k}J_{Ak}^{*}\beta_{Ak}+\epsilon\sum_{k}J_{Bk}^{*}\beta_{Bk},\\
i\Omega_{A}\frac{\partial\alpha_{A}}{\partial\tau_{A}} & = & \varepsilon_{A}^{\prime}\alpha_{A}+\epsilon\sum_{k}\xi_{Ak}^{*}\beta_{Ak}+\epsilon\sum_{k}\xi_{Bk}^{*}\beta_{Bk},\\
i\Omega_{Ak}\frac{\partial\beta_{Ak}}{\partial\tau_{Ak}} & = & \varepsilon_{k}^{-}\beta_{Ak}+\epsilon\xi_{Ak}\alpha_{A}+\epsilon J_{Ak}\alpha_{b},\\
i\Omega_{Bk}\frac{\partial\beta_{Bk}}{\partial\tau_{Bk}} & = & -\varepsilon_{k}^{+}\beta_{Bk}+\epsilon\xi_{Bk}\alpha_{A}+\epsilon J_{Bk}\alpha_{b}.
\end{eqnarray}
\end{subequations}

By inserting Eq.~(\ref{eq:RenorFre}) into Eq.~(\ref{eq:Diff}),
and expanding the coefficients as\begin{subequations}
\begin{eqnarray}
\alpha_{b} & = & \alpha_{b}^{(0)}+\epsilon\alpha_{b}^{(1)}+\epsilon^{2}\alpha_{b}^{(2)}+\epsilon^{3}\alpha_{b}^{(3)}+\ldots,\\
\alpha_{A} & = & \alpha_{A}^{(0)}+\epsilon\alpha_{A}^{(1)}+\epsilon^{2}\alpha_{A}^{(2)}+\epsilon^{3}\alpha_{A}^{(3)}+\ldots,\\
\beta_{Ak} & = & \beta_{Ak}^{(0)}+\epsilon\beta_{Ak}^{(1)}+\epsilon^{2}\beta_{Ak}^{(2)}+\epsilon^{3}\beta_{Ak}^{(3)}+\ldots,\\
\beta_{Bk} & = & \beta_{Bk}^{(0)}+\epsilon\beta_{Bk}^{(1)}+\epsilon^{2}\beta_{Bk}^{(2)}+\epsilon^{3}\beta_{Bk}^{(3)}+\ldots,
\end{eqnarray}
\end{subequations}we could obtain a set of equations for different
orders of coefficients,\begin{subequations}
\begin{align}
 & i\left(\omega+\epsilon\Omega_{b}^{(1)}+\epsilon^{2}\Omega_{b}^{(2)}+\epsilon^{3}\Omega_{b}^{(3)}\ldots\right)\frac{\partial\left(\alpha_{b}^{(0)}+\epsilon\alpha_{b}^{(1)}+\epsilon^{2}\alpha_{b}^{(2)}+\epsilon^{3}\alpha_{b}^{(3)}\ldots\right)}{\partial\tau_{b}}\nonumber \\
= & \omega\left(\alpha_{b}^{(0)}+\epsilon\alpha_{b}^{(1)}+\epsilon^{2}\alpha_{b}^{(2)}+\epsilon^{3}\alpha_{b}^{(3)}\ldots\right)+\epsilon\sum_{k}J_{Ak}^{*}\beta_{Ak}+\epsilon\sum_{k}J_{Bk}^{*}\beta_{Bk},\\
 & i\left(\varepsilon_{A}^{\prime}+\epsilon\Omega_{A}^{(1)}+\epsilon^{2}\Omega_{A}^{(2)}+\epsilon^{3}\Omega_{A}^{(3)}\ldots\right)\frac{\partial\left(\alpha_{A}^{(0)}+\epsilon\alpha_{A}^{(1)}+\epsilon^{2}\alpha_{A}^{(2)}+\epsilon^{3}\alpha_{A}^{(3)}\ldots\right)}{\partial\tau_{A}}\nonumber \\
= & \varepsilon_{A}^{\prime}\left(\alpha_{A}^{(0)}+\epsilon\alpha_{A}^{(1)}+\epsilon^{2}\alpha_{A}^{(2)}+\epsilon^{3}\alpha_{A}^{(3)}\ldots\right)+\epsilon\sum_{k}\xi_{Ak}^{*}\beta_{Ak}+\epsilon\sum_{k}\xi_{Bk}^{*}\beta_{Bk},\\
 & i\left(\varepsilon_{k}^{-}+\epsilon\Omega_{Ak}^{(1)}+\epsilon^{2}\Omega_{Ak}^{(2)}+\epsilon^{3}\Omega_{Ak}^{(3)}\ldots\right)\frac{\partial\left(\beta_{Ak}^{(0)}+\epsilon\beta_{Ak}^{(1)}+\epsilon^{2}\beta_{Ak}^{(2)}+\epsilon^{3}\beta_{Ak}^{(3)}\ldots\right)}{\partial\tau_{Ak}}\nonumber \\
= & \varepsilon_{k}^{-}\left(\beta_{Ak}^{(0)}+\epsilon\beta_{Ak}^{(1)}+\epsilon^{2}\beta_{Ak}^{(2)}+\epsilon^{3}\beta_{Ak}^{(3)}\ldots\right)+\epsilon\xi_{Ak}\alpha_{A}+\epsilon J_{Ak}\alpha_{b},\\
 & i\left(-\varepsilon_{k}^{+}+\epsilon\Omega_{Bk}^{(1)}+\epsilon^{2}\Omega_{Bk}^{(2)}+\epsilon^{3}\Omega_{Bk}^{(3)}\ldots\right)\frac{\partial\left(\beta_{Bk}^{(0)}+\epsilon\beta_{Bk}^{(1)}+\epsilon^{2}\beta_{Bk}^{(2)}+\epsilon^{3}\beta_{Bk}^{(3)}\ldots\right)}{\partial\tau_{Bk}}\nonumber \\
= & -\varepsilon_{k}^{+}\left(\beta_{Bk}^{(0)}+\epsilon\beta_{Bk}^{(1)}+\epsilon^{2}\beta_{Bk}^{(2)}+\epsilon^{3}\beta_{Bk}^{(3)}\ldots\right)+\epsilon\xi_{Bk}\alpha_{A}+\epsilon J_{Bk}\alpha_{b}.
\end{align}
\end{subequations}

For the zeroth-order coefficients, we have\begin{subequations}
\begin{eqnarray}
\omega(i\frac{\partial\alpha_{b}^{(0)}}{\partial\tau_{b}}-\alpha_{b}^{(0)}) & = & 0,\\
\varepsilon_{A}^{\prime}(i\frac{\partial\alpha_{A}^{(0)}}{\partial\tau_{A}}-\alpha_{A}^{(0)}) & = & 0,\\
\varepsilon_{k}^{-}(i\frac{\partial\beta_{Ak}^{(0)}}{\partial\tau_{Ak}}-\beta_{Ak}^{(0)}) & = & 0,\\
\varepsilon_{k}^{+}(i\frac{\partial\beta_{Bk}^{(0)}}{\partial\tau_{Bk}}-\beta_{Bk}^{(0)}) & = & 0.
\end{eqnarray}
\end{subequations}The solutions to the above equations are\begin{subequations}\label{eq:0thOrdCoe}
\begin{eqnarray}
\alpha_{b}^{(0)} & = & A_{b}e^{-i\tau_{b}},\\
\alpha_{A}^{(0)} & = & A_{A}e^{-i\tau_{A}},\\
\beta_{Ak}^{(0)} & = & B_{Ak}e^{-i\tau_{Ak}},\\
\beta_{Bk}^{(0)} & = & B_{Bk}e^{-i\tau_{Bk}},
\end{eqnarray}
\end{subequations}where the constants $A_{u}$ ($u=$b,A) and $B_{vk}$
($v=$A,B) will be determined by the initial condition later.

For the first-order coefficients, we have\begin{subequations}\label{eq:1stOrdEq}
\begin{eqnarray}
\omega(i\frac{\partial\alpha_{b}^{(1)}}{\partial\tau_{b}}-\alpha_{b}^{(1)}) & = & -i\Omega_{b}^{(1)}\frac{\partial\alpha_{b}^{(0)}}{\partial\tau_{b}}+\sum_{k}J_{Ak}^{*}\beta_{Ak}^{(0)}+\sum_{k}J_{Bk}^{*}\beta_{Bk}^{(0)},\\
\varepsilon_{A}^{\prime}(i\frac{\partial\alpha_{A}^{(1)}}{\partial\tau_{A}}-\alpha_{A}^{(1)}) & = & -i\Omega_{A}^{(1)}\frac{\partial\alpha_{A}^{(0)}}{\partial\tau_{A}}+\sum_{k}\xi_{Ak}^{*}\beta_{Ak}^{(0)}+\sum_{k}\xi_{Bk}^{*}\beta_{Bk}^{(0)},\\
\varepsilon_{k}^{-}(i\frac{\partial\beta_{Ak}^{(1)}}{\partial\tau_{Ak}}-\beta_{Ak}^{(1)}) & = & -i\Omega_{Ak}^{(1)}\frac{\partial\beta_{Ak}^{(0)}}{\partial\tau_{Ak}}+\xi_{Ak}\alpha_{A}^{(0)}+J_{Ak}\alpha_{b}^{(0)},\\
-\varepsilon_{k}^{+}(i\frac{\partial\beta_{Bk}^{(1)}}{\partial\tau_{Bk}}-\beta_{Bk}^{(1)}) & = & -i\Omega_{Bk}^{(1)}\frac{\partial\beta_{Bk}^{(0)}}{\partial\tau_{Bk}}+\xi_{Bk}\alpha_{A}^{(0)}+J_{Bk}\alpha_{b}^{(0)}.
\end{eqnarray}
\end{subequations}The first terms on the r.h.s will lead to divergence
in the long-time limit, because they are resonant driving. As a result,
the first-order renormalizations to energies vanish, i.e.
\begin{equation}
\Omega_{b}^{(1)}=\Omega_{A}^{(1)}=\Omega_{Ak}^{(1)}=\Omega_{Bk}^{(1)}=0.
\end{equation}
By further inserting Eq.~(\ref{eq:0thOrdCoe}) into Eq.~(\ref{eq:1stOrdEq}):\begin{subequations}
\begin{eqnarray}
\omega(i\frac{\partial\alpha_{b}^{(1)}}{\partial\tau_{b}}-\alpha_{b}^{(1)}) & = & \sum_{k}J_{Ak}^{*}B_{Ak}e^{-i\tau_{Ak}}+\sum_{k}J_{Bk}^{*}B_{Bk}e^{-i\tau_{Bk}},\\
\varepsilon_{A}^{\prime}(i\frac{\partial\alpha_{A}^{(1)}}{\partial\tau_{A}}-\alpha_{A}^{(1)}) & = & \sum_{k}\xi_{Ak}^{*}B_{Ak}e^{-i\tau_{Ak}}+\sum_{k}\xi_{Bk}^{*}B_{Bk}e^{-i\tau_{Bk}},\\
\varepsilon_{k}^{-}(i\frac{\partial\beta_{Ak}^{(1)}}{\partial\tau_{Ak}}-\beta_{Ak}^{(1)}) & = & \xi_{Ak}A_{A}e^{-i\tau_{A}}+J_{Ak}A_{b}e^{-i\tau_{b}},\\
-\varepsilon_{k}^{+}(i\frac{\partial\beta_{Bk}^{(1)}}{\partial\tau_{Bk}}-\beta_{Bk}^{(1)}) & = & \xi_{Bk}A_{A}e^{-i\tau_{A}}+J_{Bk}A_{b}e^{-i\tau_{b}},
\end{eqnarray}
\end{subequations}we obtain the first-order terms as\begin{subequations}
\begin{eqnarray}
\alpha_{b}^{(1)} & = & \sum_{k}B_{Ak}\frac{\Omega_{b}J_{Ak}^{*}}{\omega(\Omega_{Ak}-\Omega_{b})}e^{-i\tau_{Ak}}+\sum_{k}B_{Bk}\frac{\Omega_{b}J_{Bk}^{*}}{\omega(\Omega_{Bk}-\Omega_{b})}e^{-i\tau_{Bk}},\\
\alpha_{A}^{(1)} & = & \sum_{k}B_{Ak}\frac{\Omega_{A}\xi_{Ak}^{*}}{\varepsilon_{A}^{\prime}(\Omega_{Ak}-\Omega_{A})}e^{-i\tau_{Ak}}+\sum_{k}B_{Bk}\frac{\Omega_{A}\xi_{Bk}^{*}}{\varepsilon_{A}^{\prime}(\Omega_{Bk}-\Omega_{A})}e^{-i\tau_{Bk}},\\
\beta_{Ak}^{(1)} & = & A_{A}\frac{\Omega_{Ak}\xi_{Ak}}{\varepsilon_{k}^{-}(\Omega_{A}-\Omega_{Ak})}e^{-i\tau_{A}}+A_{b}\frac{\Omega_{Ak}J_{Ak}}{\varepsilon_{k}^{-}(\Omega_{b}-\Omega_{Ak})}e^{-i\tau_{b}},\\
\beta_{Bk}^{(1)} & = & A_{A}\frac{\Omega_{Bk}\xi_{Bk}}{-\varepsilon_{k}^{+}(\Omega_{A}-\Omega_{Bk})}e^{-i\tau_{A}}+A_{b}\frac{\Omega_{Bk}J_{Bk}}{-\varepsilon_{k}^{+}(\Omega_{b}-\Omega_{Bk})}e^{-i\tau_{b}}.
\end{eqnarray}
\end{subequations}

In the same way, we can obtain the equations for the second-order
coefficients as\begin{subequations}
\begin{eqnarray}
\omega(i\frac{\partial\alpha_{b}^{(2)}}{\partial\tau_{b}}-\alpha_{b}^{(2)}) & = & -i\Omega_{b}^{(2)}\frac{\partial\alpha_{b}^{(0)}}{\partial\tau_{b}}+\sum_{k}J_{Ak}^{*}(A_{A}\frac{\Omega_{Ak}\xi_{Ak}}{\varepsilon_{k}^{-}(\Omega_{A}-\Omega_{Ak})}e^{-i\tau_{A}}+A_{b}\frac{\Omega_{Ak}J_{Ak}}{\varepsilon_{k}^{-}(\Omega_{b}-\Omega_{Ak})}e^{-i\tau_{b}}]\nonumber \\
 &  & +\sum_{k}J_{Bk}^{*}[A_{A}\frac{\Omega_{Bk}\xi_{Bk}}{-\varepsilon_{k}^{+}(\Omega_{A}-\Omega_{Bk})}e^{-i\tau_{A}}+A_{b}\frac{\Omega_{Bk}J_{Bk}}{-\varepsilon_{k}^{+}(\Omega_{b}-\Omega_{Bk})}e^{-i\tau_{b}}],\\
\varepsilon_{A}^{\prime}(i\frac{\partial\alpha_{A}^{(2)}}{\partial\tau_{A}}-\alpha_{A}^{(2)}) & = & -i\Omega_{A}^{(2)}\frac{\partial\alpha_{A}^{(0)}}{\partial\tau_{A}}+\sum_{k}\xi_{Ak}^{*}[A_{A}\frac{\Omega_{Ak}\xi_{Ak}}{\varepsilon_{k}^{-}(\Omega_{A}-\Omega_{Ak})}e^{-i\tau_{A}}+A_{b}\frac{\Omega_{Ak}J_{Ak}}{\varepsilon_{k}^{-}(\Omega_{b}-\Omega_{Ak})}e^{-i\tau_{b}}]\nonumber \\
 &  & +\sum_{k}\xi_{Bk}^{*}[A_{A}\frac{\Omega_{Bk}\xi_{Bk}}{-\varepsilon_{k}^{+}(\Omega_{A}-\Omega_{Bk})}e^{-i\tau_{A}}+A_{b}\frac{\Omega_{Bk}J_{Bk}}{-\varepsilon_{k}^{+}(\Omega_{b}-\Omega_{Bk})}e^{-i\tau_{b}}],\\
\varepsilon_{k}^{-}(i\frac{\partial\beta_{Ak}^{(2)}}{\partial\tau_{Ak}}-\beta_{Ak}^{(2)}) & = & -i\Omega_{Ak}^{(2)}\frac{\partial\beta_{Ak}^{(0)}}{\partial\tau_{Ak}}+\xi_{Ak}[\sum_{k^{\prime}}B_{Ak^{\prime}}\frac{\Omega_{A}\xi_{Ak^{\prime}}^{*}}{\varepsilon_{A}^{\prime}(\Omega_{Ak^{\prime}}-\Omega_{A})}e^{-i\tau_{Ak^{\prime}}}+\sum_{k^{\prime}}B_{Bk^{\prime}}\frac{\Omega_{A}\xi_{Bk^{\prime}}^{*}}{\varepsilon_{A}^{\prime}(\Omega_{Bk^{\prime}}-\Omega_{A})}e^{-i\tau_{Bk^{\prime}}}]\nonumber \\
 &  & +J_{Ak}[\sum_{k^{\prime}}B_{Ak^{\prime}}\frac{\Omega_{b}J_{Ak^{\prime}}^{*}}{\omega(\Omega_{Ak^{\prime}}-\Omega_{b})}e^{-i\tau_{Ak^{\prime}}}+\sum_{k^{\prime}}B_{Bk^{\prime}}\frac{\Omega_{b}J_{Bk^{\prime}}^{*}}{\omega(\Omega_{Bk^{\prime}}-\Omega_{b})}e^{-i\tau_{Bk^{\prime}}}],\\
-\varepsilon_{k}^{+}(i\frac{\partial\beta_{Bk}^{(2)}}{\partial\tau_{Bk}}-\beta_{Bk}^{(2)}) & = & -i\Omega_{Bk}^{(2)}\frac{\partial\beta_{Bk}^{(0)}}{\partial\tau_{Bk}}+\xi_{Bk}[\sum_{k^{\prime}}B_{Ak^{\prime}}\frac{\Omega_{A}\xi_{Ak^{\prime}}^{*}}{\varepsilon_{A}^{\prime}(\Omega_{Ak^{\prime}}-\Omega_{A})}e^{-i\tau_{Ak^{\prime}}}+\sum_{k^{\prime}}B_{Bk^{\prime}}\frac{\Omega_{A}\xi_{Bk^{\prime}}^{*}}{\varepsilon_{A}^{\prime}(\Omega_{Bk^{\prime}}-\Omega_{A})}e^{-i\tau_{Bk^{\prime}}}]\nonumber \\
 &  & +J_{Bk}[\sum_{k^{\prime}}B_{Ak^{\prime}}\frac{\Omega_{b}J_{Ak^{\prime}}^{*}}{\omega(\Omega_{Ak^{\prime}}-\Omega_{b})}e^{-i\tau_{Ak^{\prime}}}+\sum_{k^{\prime}}B_{Bk^{\prime}}\frac{\Omega_{b}J_{Bk^{\prime}}^{*}}{\omega(\Omega_{Bk^{\prime}}-\Omega_{b})}e^{-i\tau_{Bk^{\prime}}}].
\end{eqnarray}
\end{subequations}Because the resonant-driving terms will result
in divergence, the second-order energies are thus\begin{subequations}
\begin{eqnarray}
\Omega_{b}^{(2)} & = & \sum_{k}[\frac{\Omega_{Ak}|J_{Ak}|^{2}}{\varepsilon_{k}^{-}(\Omega_{b}-\Omega_{Ak})}+\frac{\Omega_{Bk}|J_{Bk}|^{2}}{-\varepsilon_{k}^{+}(\Omega_{b}-\Omega_{Bk})}],\\
\Omega_{A}^{(2)} & = & \sum_{k}[\frac{\Omega_{Ak}|\xi_{Ak}|^{2}}{\varepsilon_{k}^{-}(\Omega_{A}-\Omega_{Ak})}+\frac{\Omega_{Bk}|\xi_{Bk}|^{2}}{-\varepsilon_{k}^{+}(\Omega_{A}-\Omega_{Bk})}],\\
\Omega_{Ak}^{(2)} & = & \frac{\Omega_{A}|\xi_{Ak}|^{2}}{\varepsilon_{A}^{\prime}(\Omega_{Ak}-\Omega_{A})}+\frac{\Omega_{b}|J_{Ak}|^{2}}{\omega(\Omega_{Ak}-\Omega_{b})},\\
\Omega_{Bk}^{(2)} & = & \frac{\Omega_{A}|\xi_{Bk}|^{2}}{\varepsilon_{A}^{\prime}(\Omega_{Bk}-\Omega_{A})}+\frac{\Omega_{b}|J_{Bk}|^{2}}{\omega(\Omega_{Bk}-\Omega_{b})}.
\end{eqnarray}
\end{subequations}And the corresponding second-order coefficients
are\begin{subequations}

\begin{eqnarray}
\alpha_{b}^{(2)} & = & \sum_{k}A_{A}[\frac{J_{Ak}^{*}\xi_{Ak}\Omega_{Ak}}{\varepsilon_{k}^{-}(\Omega_{A}-\Omega_{Ak})}+\frac{J_{Bk}^{*}\xi_{Bk}\Omega_{Bk}}{-\varepsilon_{k}^{+}(\Omega_{A}-\Omega_{Bk})}]\frac{\Omega_{b}e^{-i\tau_{A}}}{\omega(\Omega_{A}-\Omega_{b})},\\
\alpha_{A}^{(2)} & = & \sum_{k}A_{b}[\frac{J_{Ak}\xi_{Ak}^{*}\Omega_{Ak}}{\varepsilon_{k}^{-}(\Omega_{b}-\Omega_{Ak})}+\frac{J_{Bk}\xi_{Bk}^{*}\Omega_{Bk}}{-\varepsilon_{k}^{+}(\Omega_{b}-\Omega_{Bk})}]\frac{\Omega_{A}e^{-i\tau_{b}}}{\varepsilon_{A}^{\prime}(\Omega_{b}-\Omega_{A})},\\
\beta_{Ak}^{(2)} & = & \sum_{k^{\prime}}^{\prime}B_{Ak^{\prime}}[\frac{\Omega_{A}\xi_{Ak}\xi_{Ak^{\prime}}^{*}}{\varepsilon_{A}^{\prime}(\Omega_{Ak^{\prime}}-\Omega_{A})}+\frac{\Omega_{b}J_{Ak}J_{Ak^{\prime}}^{*}}{\omega(\Omega_{Ak^{\prime}}-\Omega_{b})}]\frac{\Omega_{Ak}e^{-i\tau_{Ak^{\prime}}}}{\varepsilon_{k}^{-}(\Omega_{Ak^{\prime}}-\Omega_{Ak})}\nonumber \\
 &  & +\sum_{k^{\prime}}B_{Bk^{\prime}}[\frac{\Omega_{A}\xi_{Ak}\xi_{Bk^{\prime}}^{*}}{\varepsilon_{A}^{\prime}(\Omega_{Bk^{\prime}}-\Omega_{A})}+\frac{\Omega_{b}J_{Ak}J_{Bk^{\prime}}^{*}}{\omega(\Omega_{Bk^{\prime}}-\Omega_{b})}]\frac{\Omega_{Ak}e^{-i\tau_{Bk^{\prime}}}}{\varepsilon_{k}^{-}(\Omega_{Bk^{\prime}}-\Omega_{Ak})},\\
\beta_{Bk}^{(2)} & = & \sum_{k^{\prime}}^{\prime}B_{Bk^{\prime}}[\frac{\Omega_{A}\xi_{Bk}\xi_{Bk^{\prime}}^{*}}{\varepsilon_{A}^{\prime}(\Omega_{Bk^{\prime}}-\Omega_{A})}+\frac{\Omega_{b}J_{Bk}J_{Bk^{\prime}}^{*}}{\omega(\Omega_{Bk^{\prime}}-\Omega_{b})}]\frac{\Omega_{Bk}e^{-i\tau_{Bk^{\prime}}}}{-\varepsilon_{k}^{+}(\Omega_{Bk^{\prime}}-\Omega_{Bk})}\nonumber \\
 &  & +\sum_{k^{\prime}}B_{Ak^{\prime}}[\frac{\Omega_{A}\xi_{Bk}\xi_{Ak^{\prime}}^{*}}{\varepsilon_{A}^{\prime}(\Omega_{Ak^{\prime}}-\Omega_{A})}+\frac{\Omega_{b}J_{Bk}J_{Ak^{\prime}}^{*}}{\omega(\Omega_{Ak^{\prime}}-\Omega_{b})}]\frac{\Omega_{Bk}e^{-i\tau_{Ak^{\prime}}}}{-\varepsilon_{k}^{+}(\Omega_{Ak^{\prime}}-\Omega_{Bk})},
\end{eqnarray}
\end{subequations}where the prime over summation $\sum_{k^{\prime}}^{\prime}$
indicates that the term related to $k^{\prime}=k$ is removed from
the summation.

For the third-order coefficients, we have\begin{subequations}

\begin{eqnarray}
i\omega\frac{\partial\alpha_{b}^{(3)}}{\partial\tau_{b}}-\omega\alpha_{b}^{(3)} & = & -i\Omega_{b}^{(2)}\frac{\partial\alpha_{b}^{(1)}}{\partial\tau_{b}}-i\Omega_{b}^{(3)}\frac{\partial\alpha_{b}^{(0)}}{\partial\tau_{b}}+\sum_{k}J_{Ak}^{*}\beta_{Ak}^{(2)}+\sum_{k}J_{Bk}^{*}\beta_{Bk}^{(2)},\\
i\varepsilon_{A}^{\prime}\frac{\partial\alpha_{A}^{(3)}}{\partial\tau_{A}}-\varepsilon_{A}^{\prime}\alpha_{A}^{(3)} & = & -i\Omega_{A}^{(2)}\frac{\partial\alpha_{A}^{(1)}}{\partial\tau_{A}}-i\Omega_{A}^{(3)}\frac{\partial\alpha_{A}^{(0)}}{\partial\tau_{A}}+\sum_{k}\xi_{Ak}^{*}\beta_{Ak}^{(2)}+\sum_{k}\xi_{Bk}^{*}\beta_{Bk}^{(2)},\\
i\varepsilon_{k}^{-}\frac{\partial\beta_{Ak}^{(3)}}{\partial\tau_{Ak}}-\varepsilon_{k}^{-}\beta_{Ak}^{(3)} & = & -i\Omega_{Ak}^{(2)}\frac{\partial\beta_{Ak}^{(1)}}{\partial\tau_{Ak}}-i\Omega_{Ak}^{(3)}\frac{\partial\beta_{Ak}^{(0)}}{\partial\tau_{Ak}}+\xi_{Ak}\sum_{k}A_{b}[\frac{J_{Ak}\xi_{Ak}^{*}\Omega_{Ak}}{\varepsilon_{k}^{-}(\Omega_{b}-\Omega_{Ak})}+\frac{J_{Bk}\xi_{Bk}^{*}\Omega_{Bk}}{-\varepsilon_{k}^{+}(\Omega_{b}-\Omega_{Bk})}]\frac{\Omega_{A}e^{-i\tau_{b}}}{\varepsilon_{A}^{\prime}(\Omega_{b}-\Omega_{A})}\nonumber \\
 &  & +J_{Ak}\sum_{k}A_{A}[\frac{J_{Ak}^{*}\xi_{Ak}\Omega_{Ak}}{\varepsilon_{k}^{-}(\Omega_{A}-\Omega_{Ak})}+\frac{J_{Bk}^{*}\xi_{Bk}\Omega_{Bk}}{-\varepsilon_{k}^{+}(\Omega_{A}-\Omega_{Bk})}]\frac{\Omega_{b}e^{-i\tau_{A}}}{\omega(\Omega_{A}-\Omega_{b})},\\
-i\varepsilon_{k}^{+}\frac{\partial\beta_{Bk}^{(3)}}{\partial\tau_{Bk}}+\varepsilon_{k}^{+}\beta_{Bk}^{(3)} & = & -i\Omega_{Bk}^{(2)}\frac{\partial\beta_{Bk}^{(1)}}{\partial\tau_{Bk}}-i\Omega_{Bk}^{(3)}\frac{\partial\beta_{Bk}^{(0)}}{\partial\tau_{Bk}}+\xi_{Bk}\sum_{k}A_{b}[\frac{J_{Ak}\xi_{Ak}^{*}\Omega_{Ak}}{\varepsilon_{k}^{-}(\Omega_{b}-\Omega_{Ak})}+\frac{J_{Bk}\xi_{Bk}^{*}\Omega_{Bk}}{-\varepsilon_{k}^{+}(\Omega_{b}-\Omega_{Bk})}]\frac{\Omega_{A}e^{-i\tau_{b}}}{\varepsilon_{A}^{\prime}(\Omega_{b}-\Omega_{A})}\nonumber \\
 &  & +J_{Bk}\sum_{k}A_{A}[\frac{J_{Ak}^{*}\xi_{Ak}\Omega_{Ak}}{\varepsilon_{k}^{-}(\Omega_{A}-\Omega_{Ak})}+\frac{J_{Bk}^{*}\xi_{Bk}\Omega_{Bk}}{-\varepsilon_{k}^{+}(\Omega_{A}-\Omega_{Bk})}]\frac{\Omega_{b}e^{-i\tau_{A}}}{\omega(\Omega_{A}-\Omega_{b})}.
\end{eqnarray}
\end{subequations}

The coefficients of resonant terms are zero and thus lead to
\begin{eqnarray}
\Omega_{b}^{(3)} & = & \Omega_{A}^{(3)}=\Omega_{Ak}^{(3)}=\Omega_{Bk}^{(3)}=0.
\end{eqnarray}

In conclusion, to the order of $\epsilon^{3}$, the renormalized energies
are\begin{subequations}
\begin{eqnarray}
\Omega_{b} & = & \omega+\epsilon^{2}\sum_{k}(\frac{|J_{Ak}|^{2}}{\omega-\varepsilon_{k}^{-}}+\frac{|J_{Bk}|^{2}}{\omega+\varepsilon_{k}^{+}}),\\
\Omega_{A} & = & \varepsilon_{A}^{\prime}+\epsilon^{2}\sum_{k}(\frac{|\xi_{Ak}|^{2}}{\varepsilon_{A}^{\prime}-\varepsilon_{k}^{-}}+\frac{|\xi_{Bk}|^{2}}{\varepsilon_{A}^{\prime}+\varepsilon_{k}^{+}}),\\
\Omega_{Ak} & = & \varepsilon_{k}^{-}+\epsilon^{2}(\frac{|\xi_{Ak}|^{2}}{\varepsilon_{k}^{-}-\varepsilon_{A}^{\prime}}+\frac{|J_{Ak}|^{2}}{\varepsilon_{k}^{-}-\omega}),\\
\Omega_{Bk} & = & -\varepsilon_{k}^{+}+\epsilon^{2}(\frac{|\xi_{Bk}|^{2}}{-\varepsilon_{k}^{+}-\varepsilon_{A}^{\prime}}+\frac{|J_{Bk}|^{2}}{-\varepsilon_{k}^{+}-\omega}).
\end{eqnarray}
\end{subequations}To the order of $\epsilon^{2}$, the coefficients
are\begin{subequations}

\begin{eqnarray}
\alpha_{b}(t) & = & A_{b}e^{-i\tau_{b}}+\epsilon\sum_{k}[B_{Ak}\frac{\Omega_{b}J_{Ak}^{*}}{\omega(\Omega_{Ak}-\Omega_{b})}e^{-i\tau_{Ak}}+B_{Bk}\frac{\Omega_{b}J_{Bk}^{*}}{\omega(\Omega_{Bk}-\Omega_{b})}e^{-i\tau_{Bk}}]\nonumber \\
 &  & +\epsilon^{2}\sum_{k}A_{A}[\frac{J_{Ak}^{*}\xi_{Ak}\Omega_{Ak}}{\varepsilon_{k}^{-}(\Omega_{A}-\Omega_{Ak})}+\frac{J_{Bk}^{*}\xi_{Bk}\Omega_{Bk}}{-\varepsilon_{k}^{+}(\Omega_{A}-\Omega_{Bk})}]\frac{\Omega_{b}e^{-i\tau_{A}}}{\omega(\Omega_{A}-\Omega_{b})},\\
\alpha_{A}(t) & = & A_{A}e^{-i\tau_{A}}+\epsilon\sum_{k}[B_{Ak}\frac{\Omega_{A}\xi_{Ak}^{*}}{\varepsilon_{A}^{\prime}(\Omega_{Ak}-\Omega_{A})}e^{-i\tau_{Ak}}+B_{Bk}\frac{\Omega_{A}\xi_{Bk}^{*}}{\varepsilon_{A}^{\prime}(\Omega_{Bk}-\Omega_{A})}e^{-i\tau_{Bk}}]\nonumber \\
 &  & +\epsilon^{2}\sum_{k}A_{b}[\frac{J_{Ak}\xi_{Ak}^{*}\Omega_{Ak}}{\varepsilon_{k}^{-}(\Omega_{b}-\Omega_{Ak})}+\frac{J_{Bk}\xi_{Bk}^{*}\Omega_{Bk}}{-\varepsilon_{k}^{+}(\Omega_{b}-\Omega_{Bk})}]\frac{\Omega_{A}e^{-i\tau_{b}}}{\varepsilon_{A}^{\prime}(\Omega_{b}-\Omega_{A})},\\
\beta_{Ak}(t) & = & B_{Ak}e^{-i\tau_{Ak}}+\epsilon[A_{A}\frac{\Omega_{Ak}\xi_{Ak}}{\varepsilon_{k}^{-}(\Omega_{A}-\Omega_{Ak})}e^{-i\tau_{A}}+A_{b}\frac{\Omega_{Ak}J_{Ak}}{\varepsilon_{k}^{-}(\Omega_{b}-\Omega_{Ak})}e^{-i\tau_{b}}]\nonumber \\
 &  & +\epsilon^{2}\{\sum_{k^{\prime}}^{\prime}B_{Ak^{\prime}}[\frac{\Omega_{A}\xi_{Ak}\xi_{Ak^{\prime}}^{*}}{\varepsilon_{A}^{\prime}(\Omega_{Ak^{\prime}}-\Omega_{A})}+\frac{\Omega_{b}J_{Ak}J_{Ak^{\prime}}^{*}}{\omega(\Omega_{Ak^{\prime}}-\Omega_{b})}]\frac{\Omega_{Ak}e^{-i\tau_{Ak^{\prime}}}}{\varepsilon_{k}^{-}(\Omega_{Ak^{\prime}}-\Omega_{Ak})}\nonumber \\
 &  & +\sum_{k^{\prime}}B_{Bk^{\prime}}[\frac{\Omega_{A}\xi_{Ak}\xi_{Bk^{\prime}}^{*}}{\varepsilon_{A}^{\prime}(\Omega_{Bk^{\prime}}-\Omega_{A})}+\frac{\Omega_{b}J_{Ak}J_{Bk^{\prime}}^{*}}{\omega(\Omega_{Bk^{\prime}}-\Omega_{b})}]\frac{\Omega_{Ak}e^{-i\tau_{Bk^{\prime}}}}{\varepsilon_{k}^{-}(\Omega_{Bk^{\prime}}-\Omega_{Ak})}\},\\
\beta_{Bk}(t) & = & B_{Bk}e^{-i\tau_{Bk}}+\epsilon[A_{A}\frac{\Omega_{Bk}\xi_{Bk}}{-\varepsilon_{k}^{+}(\Omega_{A}-\Omega_{Bk})}e^{-i\tau_{A}}+A_{b}\frac{\Omega_{Bk}J_{Bk}}{-\varepsilon_{k}^{+}(\Omega_{b}-\Omega_{Bk})}e^{-i\tau_{b}}]\nonumber \\
 &  & +\epsilon^{2}\{\sum_{k^{\prime}}^{\prime}B_{Bk^{\prime}}[\frac{\Omega_{A}\xi_{Bk}\xi_{Bk^{\prime}}^{*}}{\varepsilon_{A}^{\prime}(\Omega_{Bk^{\prime}}-\Omega_{A})}+\frac{\Omega_{b}J_{Bk}J_{Bk^{\prime}}^{*}}{\omega(\Omega_{Bk^{\prime}}-\Omega_{b})}]\frac{\Omega_{Bk}e^{-i\tau_{Bk^{\prime}}}}{-\varepsilon_{k}^{+}(\Omega_{Bk^{\prime}}-\Omega_{Bk})}\nonumber \\
 &  & +\sum_{k^{\prime}}B_{Ak^{\prime}}[\frac{\Omega_{A}\xi_{Bk}\xi_{Ak^{\prime}}^{*}}{\varepsilon_{A}^{\prime}(\Omega_{Ak^{\prime}}-\Omega_{A})}+\frac{\Omega_{b}J_{Bk}J_{Ak^{\prime}}^{*}}{\omega(\Omega_{Ak^{\prime}}-\Omega_{b})}]\frac{\Omega_{Bk}e^{-i\tau_{Ak^{\prime}}}}{-\varepsilon_{k}^{+}(\Omega_{Ak^{\prime}}-\Omega_{Bk})}\}.
\end{eqnarray}
\end{subequations}

By using the initial condition
\begin{eqnarray}
\alpha_{b}(0) & = & 1,\:\alpha_{A}(0)=\beta_{Ak}(0)=\beta_{Bk}(0)=0,
\end{eqnarray}
we can obtain the constants as\begin{subequations}

\begin{eqnarray}
A_{b} & = & 1-\epsilon^{2}\sum_{k}[\frac{|J_{Ak}|^{2}}{(\Omega_{b}-\Omega_{Ak})^{2}}+\frac{|J_{Bk}|^{2}}{(\Omega_{b}-\Omega_{Bk})^{2}}]\nonumber \\
 & = & 1-\sum_{k}[\frac{|J_{Ak}|^{2}}{(\omega-\varepsilon_{k}^{-})^{2}}+\frac{|J_{Bk}|^{2}}{(\omega+\varepsilon_{k}^{+})^{2}}],\\
A_{A} & = & -\epsilon^{2}\sum_{k}[\frac{J_{Ak}}{(\Omega_{b}-\Omega_{Ak})}\frac{\xi_{Ak}^{*}}{(\Omega_{A}-\Omega_{Ak})}+\frac{J_{Bk}}{(\Omega_{b}-\Omega_{Bk})}\frac{\xi_{Bk}^{*}}{(\Omega_{A}-\Omega_{Bk})}]\nonumber \\
 &  & -\epsilon^{2}\sum_{k}(\frac{J_{Ak}\xi_{Ak}^{*}}{\Omega_{b}-\Omega_{Ak}}+\frac{J_{Bk}\xi_{Bk}^{*}}{\Omega_{b}-\Omega_{Bk}})\frac{1}{\Omega_{b}-\Omega_{A}}\nonumber \\
 & = & -\sum_{k}[\frac{J_{Ak}}{(\omega-\varepsilon_{k}^{-})}\frac{\xi_{Ak}^{*}}{(\varepsilon_{A}^{\prime}-\varepsilon_{k}^{-})}+\frac{J_{Bk}}{(\omega+\varepsilon_{k}^{+})}\frac{\xi_{Bk}^{*}}{(\varepsilon_{A}^{\prime}+\varepsilon_{k}^{+})}]\nonumber \\
 &  & -\sum_{k}(\frac{J_{Ak}\xi_{Ak}^{*}}{\omega-\varepsilon_{k}^{-}}+\frac{J_{Bk}\xi_{Bk}^{*}}{\omega+\varepsilon_{k}^{+}})\frac{1}{\omega-\varepsilon_{A}^{\prime}},\\
B_{Ak} & = & -\epsilon\frac{J_{Ak}}{\Omega_{b}-\Omega_{Ak}}\nonumber \\
 & = & -\frac{J_{Ak}}{\omega-\varepsilon_{k}^{-}},\\
B_{Bk} & = & -\epsilon\frac{J_{Bk}}{\Omega_{b}-\Omega_{Bk}}\nonumber \\
 & = & -\frac{J_{Bk}}{\omega+\varepsilon_{k}^{+}}.
\end{eqnarray}
\end{subequations}

To the order of $\epsilon^{2}$ , the probability amplitudes are

\begin{eqnarray}
\alpha_{b}(t) & = & A_{b}e^{-i\tau_{b}}+\sum_{k}[B_{Ak}\frac{\Omega_{b}J_{Ak}^{*}}{\omega(\Omega_{Ak}-\Omega_{b})}e^{-i\tau_{Ak}}+B_{Bk}\frac{\Omega_{b}J_{Bk}^{*}}{\omega(\Omega_{Bk}-\Omega_{b})}e^{-i\tau_{Bk}}]\nonumber \\
 & = & A_{b}e^{-i\tau_{b}}+\sum_{k}(\frac{B_{Ak}J_{Ak}^{*}}{\varepsilon_{k}^{-}-\omega}e^{-i\tau_{Ak}}+\frac{B_{Bk}J_{Bk}^{*}}{-\varepsilon_{k}^{+}-\omega}e^{-i\tau_{Bk}}),\\
\alpha_{A}(t) & = & A_{A}e^{-i\tau_{A}}+\sum_{k}[B_{Ak}\frac{\Omega_{A}\xi_{Ak}^{*}}{\varepsilon_{A}^{\prime}(\Omega_{Ak}-\Omega_{A})}e^{-i\tau_{Ak}}+B_{Bk}\frac{\Omega_{A}\xi_{Bk}^{*}}{\varepsilon_{A}^{\prime}(\Omega_{Bk}-\Omega_{A})}e^{-i\tau_{Bk}}]\nonumber \\
 &  & +\sum_{k}A_{b}[\frac{J_{Ak}\xi_{Ak}^{*}\Omega_{Ak}}{\varepsilon_{k}^{-}(\Omega_{b}-\Omega_{Ak})}+\frac{J_{Bk}\xi_{Bk}^{*}\Omega_{Bk}}{-\varepsilon_{k}^{+}(\Omega_{b}-\Omega_{Bk})}]\frac{\Omega_{A}e^{-i\tau_{b}}}{\varepsilon_{A}^{\prime}(\Omega_{b}-\Omega_{A})}\nonumber \\
 & = & A_{A}e^{-i\tau_{A}}+\sum_{k}(\frac{B_{Ak}\xi_{Ak}^{*}}{\varepsilon_{k}^{-}-\varepsilon_{A}^{\prime}}e^{-i\tau_{Ak}}+\frac{B_{Bk}\xi_{Bk}^{*}}{-\varepsilon_{k}^{+}-\varepsilon_{A}^{\prime}}e^{-i\tau_{Bk}})\nonumber \\
 &  & +A_{b}\frac{e^{-i\tau_{b}}}{\omega-\varepsilon_{A}^{\prime}}\sum_{k}(\frac{J_{Ak}\xi_{Ak}^{*}}{\omega-\varepsilon_{k}^{-}}+\frac{J_{Bk}\xi_{Bk}^{*}}{\omega+\varepsilon_{k}^{+}}),\\
\beta_{Ak}(t) & = & B_{Ak}e^{-i\tau_{Ak}}+A_{b}\frac{\Omega_{Ak}J_{Ak}}{\varepsilon_{k}^{-}(\Omega_{b}-\Omega_{Ak})}e^{-i\tau_{b}}\nonumber \\
 & = & B_{Ak}e^{-i\tau_{Ak}}+A_{b}\frac{J_{Ak}}{\omega-\varepsilon_{k}^{-}}e^{-i\tau_{b}},\\
\beta_{Bk}(t) & = & B_{Bk}e^{-i\tau_{Bk}}+A_{b}\frac{\Omega_{Bk}J_{Bk}}{-\varepsilon_{k}^{+}(\Omega_{b}-\Omega_{Bk})}e^{-i\tau_{b}}\nonumber \\
 & = & B_{Bk}e^{-i\tau_{Bk}}+A_{b}\frac{J_{Bk}}{\omega+\varepsilon_{k}^{+}}e^{-i\tau_{b}},
\end{eqnarray}
where we have taken $\epsilon=1$.\end{widetext}

To the lowest order of $\epsilon$,
\begin{align*}
\alpha_{b}(t) & =A_{b}e^{-i\tau_{b}}\\
 & =\exp\left[-i\omega t-it\sum_{k}(\frac{|J_{Ak}|^{2}}{\omega-\varepsilon_{k}^{-}}+\frac{|J_{Bk}|^{2}}{\omega+\varepsilon_{k}^{+}})\right].
\end{align*}
Regardless of fluorescence in the donor ring, and assuming there is
a small imaginary part in $\omega$, we recover the result under Wigner-Weisskopf approximation, i.e.
\begin{align}
\alpha_{b}(t) & =A_{b}e^{-i\tau_{b}}\nonumber \\
 & =\exp\left[-i\omega t-it\sum_{k}(\frac{|J_{Ak}|^{2}}{\omega-\varepsilon_{k}+i0^{+}}+\frac{|J_{Bk}|^{2}}{\omega+\varepsilon_{k}+i0^{+}})\right]\nonumber \\
 & =\exp\left[-i(\omega+\Delta^{\prime})t-\gamma t\right],
\end{align}
where\begin{subequations}
\begin{align}
\Delta^{\prime} & =\sum_{k}\wp(\frac{|J_{Ak}|^{2}}{\omega-\varepsilon_{k}}+\frac{|J_{Bk}|^{2}}{\omega+\varepsilon_{k}}),\\
\gamma & =\pi\sum_{k}[J_{Ak}|^{2}\delta(\omega-\varepsilon_{k})+|J_{Bk}|^{2}\delta(\omega+\varepsilon_{k})].
\end{align}
\end{subequations}

\end{document}